\RequirePackage{amsmath}
\documentclass[12pt]{iopart}
\usepackage{enumerate}
\usepackage{amsmath}
\pdfoutput=1
\usepackage[paperwidth=210mm,paperheight=297mm,centering,hmargin=2cm,vmargin=2.5cm]{geometry}
\usepackage[pdftex]{graphicx}
\usepackage[colorlinks=true,citecolor=blue]{hyperref}
\usepackage{cite}
\usepackage{mathrsfs}
\usepackage{array}
\usepackage{hyperref}
\usepackage{color}
\usepackage{graphicx,epsfig}
\usepackage{amssymb}
\usepackage[normalem]{ulem}

\def\be{\begin{eqnarray}}
\def\ee{\end{eqnarray}}

\usepackage{xcolor}
\usepackage{soul}
\setstcolor{red}

\begin{document}

\title{Spin-flip Enhanced Thermoelectricity in Superconductor-Ferromagnet Bilayers}

\author{A. Rezaei, A. Kamra, P. Machon, and W. Belzig}

\address{Department of Physics, University of Konstanz, D-78457 Konstanz, Germany}
\eads{ali.rezaei@uni-konstanz.de}

\date{\today}

\begin{abstract}
We study the effects of spin-splitting and spin-flip scattering in a
superconductor (S) on the thermoelectric properties of a tunneling
contact to a metallic ferromagnet (F) using the Green's function method. 
A giant thermopower has been theoretically predicted and experimentally
observed in such structures. This is attributed to the spin-dependent 
particle-hole asymmetry in the tunneling density of states in the S/F 
heterostructure. Here, we evaluate the S density of states and thermopower 
for a range of temperatures, Zeeman-splitting, and spin-flip scattering. 
In contrast to the naive expectation based on the negative effect of 
spin-flip scattering on Cooper pairing, we find that the spin-flip scattering
strongly enhances the thermoelectric performance of the system in the
low-field and low-temperature regime. This is attributed to a complex
interplay between the charge and spin conductances caused by the
softening of the spin-dependent superconducting gaps. The maximal value 
of the thermopower exceeds $k_{B}/e$ by a factor of $\approx$ 5 and has a
nonmonotonic dependence on spin-splitting and spin-flip rate.
\end{abstract}
\maketitle
\section{Introduction} \label{sec:intro}
The field of thermoelectrics has drawn an enormous amount of interest
in recent years driven by the goal of recovering unused heat and
converting it into usable electricity. The thermoelectric (TE) or
Seebeck effect refers to the conversion of a temperature difference
($\Delta T$) across a system into the usable electrical potential
($V$). The effect is quantified in terms of the so-called Seebeck
coefficient $S$ as $S= - V/\Delta T$. The efficiency of a TE device is
determined by the type of materials used in making the device. Thus, a
major focus of efforts towards addressing today's energy challenge is
to find more efficient TE structures \cite{Disalvo1999}. 

In a conductor, an applied temperature gradient drives the flow of
electrons and holes. These two contributions are in general unequal
due to the density of states (DOS) variations vs. quasiparticle energy
giving rise to a net charge current. Under open circuit conditions, 
the requirement of no net current leads to a charge buildup that appears 
as a voltage called thermopower. However, the variation in the DOS is 
typically very small and, consequently, the TE response 
is negligibly small. Hence, achieving a strong variation in the DOS near 
the Fermi energy is the key to enhancing TE effects in a system. 

A superconductor with spin-splitting breaks the spin-dependent electron-hole 
symmetry. Its interface with a ferromagnetic metal results in a spin-polarized 
conductance leading to a large Seebeck effect. Following this principle, 
large spin-dependent TE effects have been theoretically proposed
\cite{Kalenkov2012,Machon2013,Machon2014,Giazotto2014,Kalenkov2014,
Ozaeta2014,Kalenkov2015,Giazotto2015} and experimentally demonstrated
\cite{Kolenda2016PRL,Kolenda2016BJN,Kolenda2017} in spin-split and -polarized
superconducting tunnel contacts~\cite{Hubler2012}. 
Furthermore, TE properties in a variety of systems have been investigated \cite{Hwang2016,Dutta2017,WeiPing2017}.
In most of the cases, the Zeeman-split superconducting heterostructures 
exhibiting sizeable TE effects require a large externally applied magnetic field
\cite{Kolenda2016PRL,Tedrow1971}. A recent experiment
\cite{Kolenda2017} reported a large Seebeck coefficient employing
the spin-splitting field provided by the proximity to a ferromagnetic 
insulator, instead of an applied magnetic field. The spin-orbit 
interaction has been found to reduce the TE performance 
of such structures~\cite{Bergeret2017}. The usability of these devices 
as thermal detectors has also been discussed~\cite{Heikkilä2017}.

Similar to the influence of the spin-orbit interaction~\cite{Bergeret2017}, 
spin-flip scattering may be expected to reduce the TE 
performance of such devices. This is because spin-flip scattering is 
known to have a detrimental effect on superconductivity via breaking 
of the Cooper pairs resulting in a reduction of the gap. One may 
assume a different perspective and argue that the reduction of 
the superconducting gap might enable quasiparticles at lower 
energies to contribute towards the Seebeck effect, thereby enhancing 
it. Thus, competing mechanisms are in play and a clear winner may 
not be ascertained apriori. Earlier works have investigated the 
influence of spin-flip on the critical current \cite{Faure2006} and critical
temperature \cite{Faure2006,Velez1999}, the spatial and energy
dependences of the DOS~\cite{Gusakova2006,Volkov1996,Belzig1996,Belzig1999}, 
the anomalous Green's function \cite{Linder2007}, and quasiparticles
distribution~\cite{Silaev2015,Virtanen2016} in a spin-split S. 

\begin{figure*}[t]
	\begin{centering}
		\includegraphics[width=0.8\textwidth]{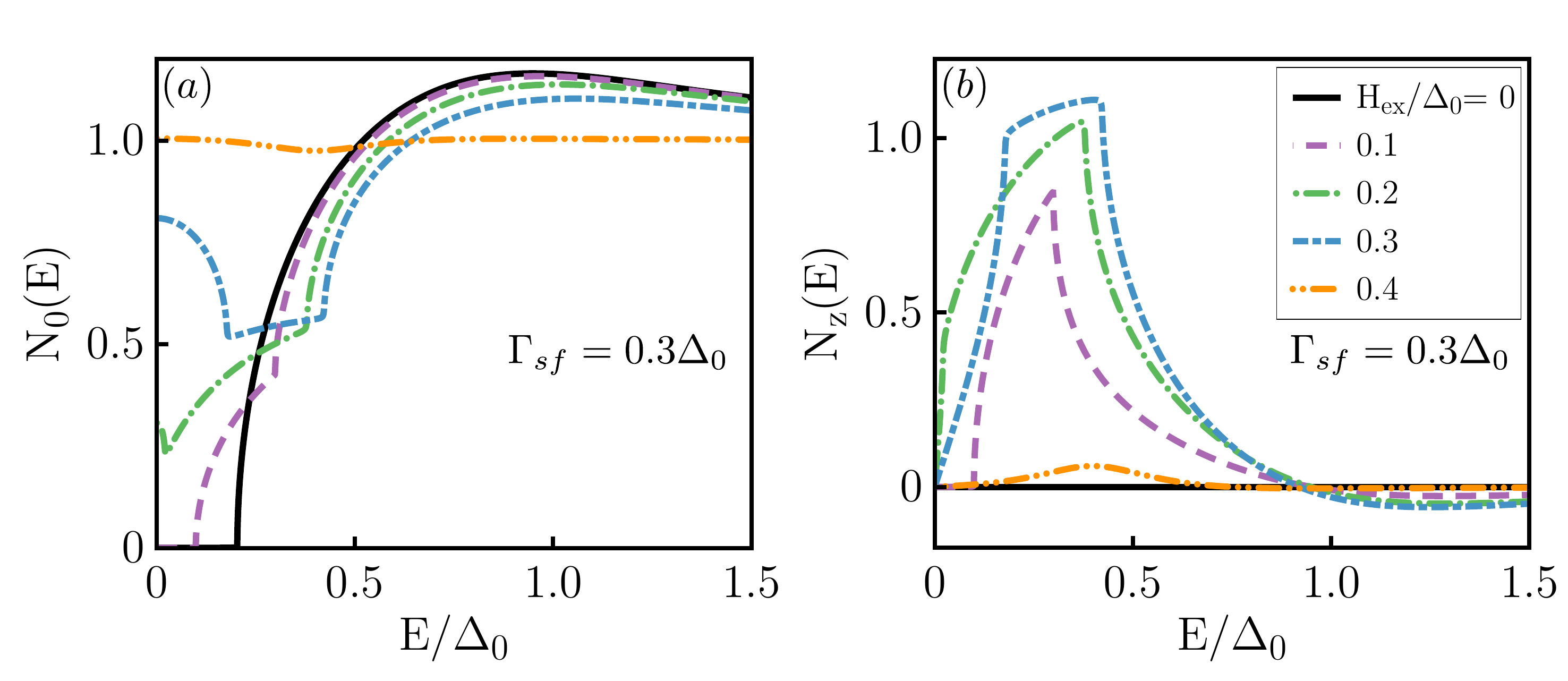}
		\caption{(Color online). Self-consistently determined
		total [$N_{0}(E)$ (left panel)] and spin-polarized [$N_{z}(E)$ 
		(right panel)] quasiparticle DOS. We employ $\Gamma_{sf}=0.3\Delta_{0}$, $k_{B}T/\Delta_{0}=0.1$ and  incoherent broadening parameter $\delta=10^{-3}\Delta_{0}$. $\Delta_0$ is the
	superconducting pair potential at $T=0$, $H_{ex}=0$ and $\Gamma_{sf}=0$}.
		\label{fig:DoS}
	\end{centering}
\end{figure*}

In this work, we theoretically address the effect of spin-flip scattering 
on the Seebeck coefficient in such S/F bilayers. We find that spin-flip 
enhances the TE response at low values of the spin-splitting. 
We incorporate spin-flip scattering and incoherent broadening self-consistently 
within the quasiclassical description of S \cite{Eilenberger1968,Belzig1999}. 
The spin-flip scattering shifts the peaks of the DOS toward zero energy [see
Fig. \ref{fig:DoS}] thus reducing the gap and eventually leading to
gapless superconductivity. We find that spin-flip scattering suppresses 
(enhances) the Seebeck coefficient at large (low) spin-splitting of the DOS. 
We find that the maximal Seebeck coefficient is relatively insensitive to 
the spin-flip rate, while the value of spin-splitting at which this maximum 
is achieved is reduced by increasing the spin-flip rate. Thus, for weak spin-flip 
scattering, its major effect is to decrease the spin-splitting required to achieve 
a large Seebeck coefficient, corresponding to the reduction of the superconducting 
gap. However, the TE response is substantial even in the gapless regime 
and starts to diminish rapidly as the superconductivity is destroyed by an increasing 
spin-flip rate. We find an analogous behavior of the figure of merit ZT.

This paper is organised as follows. In section \ref{sec:model}, we
outline the model and theoretical framework employed to describe the
TE effects in an S/F bilayer. We present and discuss our results regarding the
dependence of the Seebeck coefficient on the spin-flip scattering rate in section \ref{sec:results}. We conclude by 
summarising our findings in section \ref{sec:conclusions}.

\section{Model} \label{sec:model}
\begin{figure}[t]
	\begin{center}
		\includegraphics[width=0.8\textwidth]{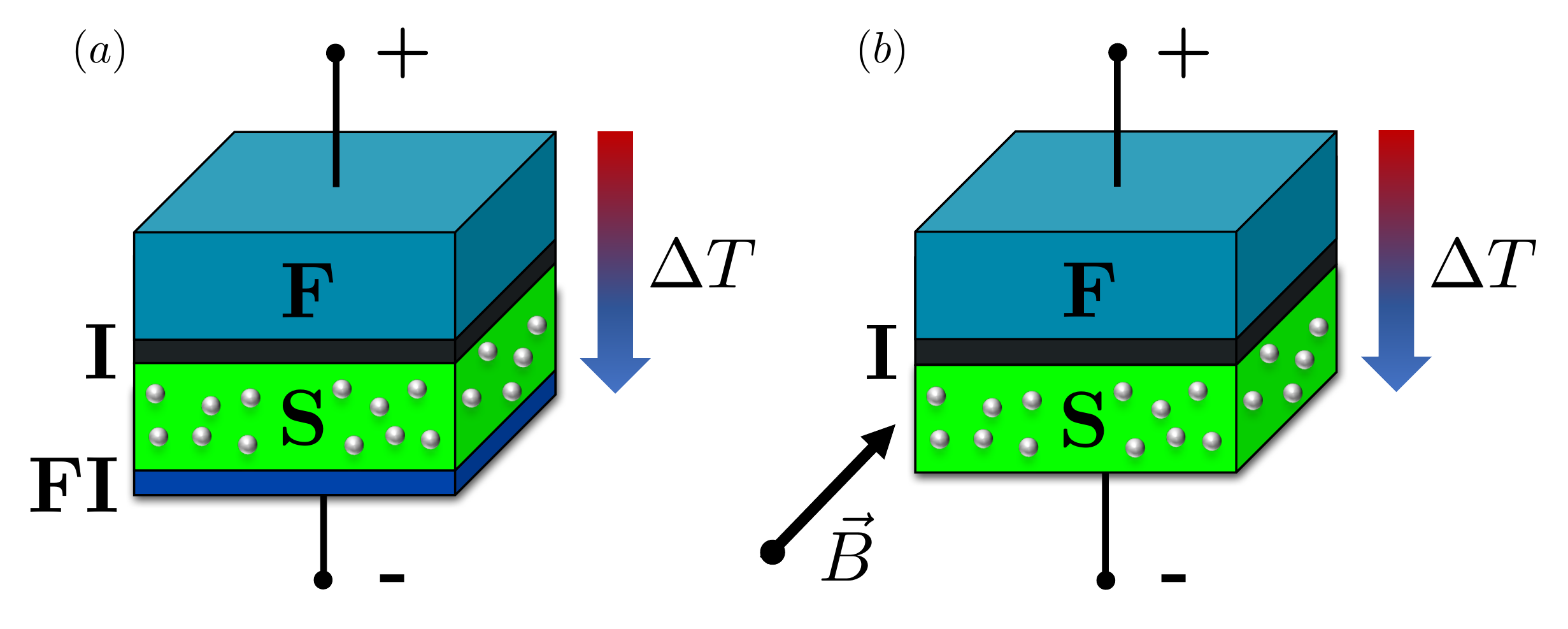}
		\caption{(Color online). Schematic illustration of the investigated 
			thin-film heterostructure. The structure consists of a ferromagnet
			(top region in light-blue) coupled to a superconductor (bottom 
			region in green) through a thin insulating barrier (black region in the center). 
			The quasiparticle DOS in the superconductor is modified by the presence
			of a spin-splitting field which may be realised via either (a)  
			proximity with a ferromagnetic insulator (FI) film (dark-blue region), 
			or (b) an applied external magnetic field. $\Delta T$ denotes the 
			temperature gradient through the junction. The presence of magnetic 
			impurities inside the superconductor or scattering at its interface 
			with the magnetic insulator results in spin-flip processes.}
		\label{fig:Schematic}
	\end{center}
\end{figure}
We consider a ferromagnetic metal (F) coupled via a spin-polarized 
tunneling contact to a spin-split superconductor (S) layer. A
sketch of the device under consideration has been shown in figure
\ref{fig:Schematic}. The spin-splitting in S may be caused by proximity 
effect with a ferromagnetic insulator
\cite{Tokuyasu1988,Hao1990,Huertas2002,Xiong2011,Bin2013} [
Fig. \ref{fig:Schematic}(a)] or an applied magnetic field ($\vec{B}$)
\cite{Tedrow1971,Catelani2008} [Fig. \ref{fig:Schematic}(b)].
The spin-splitting field is oriented collinear to the magnetization 
direction of F. The thickness of the superconducting 
film is assumed smaller than the superconducting coherence length.
The coupling between the two layers is assumed to be weak 
enough to not affect each others equilibrium properties. The 
exchange-splitting field $H_{ex}$ in the S layer acts on the spins 
of the electrons and breaks the particle-hole symmetry in the 
spin-resolved DOS. Considering the drives to be small, one 
can write TE coefficients in terms of the linear-response matrices. 

Adopting the notation of Ref.~\cite{Ozaeta2014}, we have for the 
charge $I_C$ and energy $I_E$ currents 
\begin{align}
    \begin{pmatrix} I_C \\ I_E  \end{pmatrix} = \begin{pmatrix} G &
     P \alpha  \\ P \alpha & G_{th} T  \end{pmatrix} \begin{pmatrix}
      V \\ \Delta T/T \end{pmatrix} ,\label{eqn_response1}
\end{align}
and for spin $I_S$ and `spin-energy' $ I_{E_S}$ currents
flowing through the tunnel contact 
\begin{align}
   \begin{pmatrix}  I_S \\ I_{E_S} \end{pmatrix} = \begin{pmatrix} 
   P G & \alpha \\ \alpha & P G_{th} T \end{pmatrix} \begin{pmatrix}
    V \\ \Delta T/T \end{pmatrix} ,\label{eqn_response2}
\end{align}
where $G$, $\alpha$, $P$, $G_{th}$, and $T$ are the charge
conductance, TE coefficient, spin-polarization of the tunnel 
contact, thermal conductance, and the absolute temperature 
of the system, respectively. The different coefficients may 
further be expressed in term of the spin-resolved DOS: \cite{Ozaeta2014}
\begin{subequations}
   \label{eqn_thermocoefficients}
   \begin{align}
    G&=G_T \int_{-\infty}^\infty dE N_0(E)\delta_T(E)
    \,,
    \\
    G_{th}&=\frac{G_T}{e^{2}T} \int_{-\infty}^\infty dE N_0(E) E^2 \delta_T(E)
    \,,
    \\
    \alpha&=\frac{G_T}{2e} \int_{-\infty}^\infty dE N_z(E) E \delta_T(E)
    \,.
    \end{align}
\end{subequations}
Here, $G_T$ is the conductance of the junction in the normal state, $E$ 
is the quasiparticle energy, $k_{B}$ is the Boltzmann constant, and 
$\delta_T(E)\equiv[{4 k_B T\cosh^2\left(E/2k_B T\right)}]^{-1}$ being the 
normalized difference of the quasiparticle distributions functions of 
the F and S expanded with respect to a small temperature difference 
$\Delta T$ across the junction. The total and the spin-polarized DOS in 
S $N_0(E)$ and $N_z(E)$, respectively, are defined as 
\begin{align}
    \label{eqn_N0Nz}
    N_0(E) &= (N_\uparrow(E) + N_\downarrow(E))/2  
    \,, \notag
    \\
    N_z(E)  &= N_\uparrow(E) - N_\downarrow(E)
    \,,
\end{align} 
where $N_{\uparrow (\downarrow)}$ is the DOS for spin-up 
(down) quasiparticles in S. These are evaluated employing the 
quasiclassical Green's function method as described below. 

The charge Seebeck coefficient describes the buildup of a TE voltage 
caused by a temperature gradient throughout a material as induced 
by the Seebeck effect when one opens the circuit and the net 
electric current vanishes such that $I_C=0$ \cite{Ozaeta2014}. 
Employing Eqs. (\ref{eqn_thermocoefficients}), the charge 
Seebeck coefficient is obtained as $S=-P\alpha/(GT)$. Thus, an
increase of the TE coefficient and a reduction of the charge conductance 
are required to enhance the thermopower. It is crucial to note that, a
non-zero spin-polarization of the interface is necessary to observe
the TE phenomena in the present system. From
Eqs.~\eqref{eqn_thermocoefficients}, we see that the charge
conductance is proportional to the averaged tunneling DOS $N_{0}(E)$. 
However, the TE coefficient is a function of the spin-polarized DOS 
$N_{z}(E)$. In the system under investigation, the spin Seebeck effect\cite{Uchida2008N,Uchida2010SSC,Uchida2010NM,Adachi2013,
Jaworski2010,Bosu2011} is determined by the same parameters 
as the charge Seebeck coefficient and refers to the creation of spin 
voltage due to a temperature difference. In other words, the spin 
thermopower is a way to quantify thermally produced spin voltages \cite{sunyong2016}. 
In the linear response regime, the efficiency of the device to produce 
TE power is quantified by the so-called dimensionless figure of merit 
$ZT$: $ZT=S^{2}GT/\tilde{G}_{th}$, where $\tilde{G}_{th}=G_{th}-(P\alpha)^{2}/GT$
is the energy conductance at zero current \cite{Ozaeta2014}. The larger 
the value of the $ZT$ the more efficient is the TE device \cite{Littman1961}. 
\begin{figure}[t]
	\begin{centering}
		\includegraphics[width=0.8\textwidth]{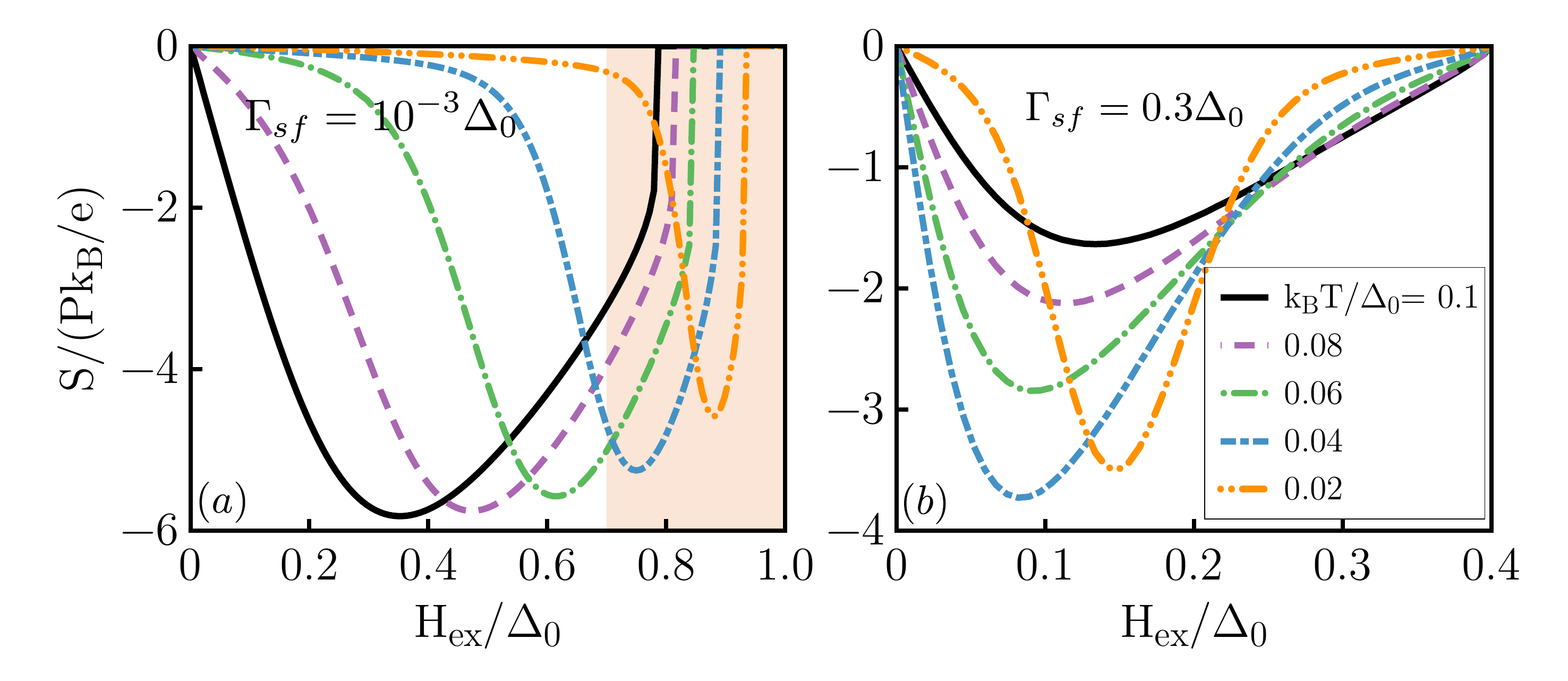}
		\caption{(Color online). Self-consistently determined
			Seebeck coefficient $S$ normalized to $Pk_{B}/e$ vs. normalized 
			spin-splitting field $H_{ex}/\Delta_{0}$. The curves are shown for 
			different temperatures $k_{B}T/\Delta_{0}=0.1$ (black solid line), $0.08$
			(purple dashed line), $0.06$ (green dash-dot line), $0.04$ (blue
			long dash-dash) and $0.02$ (orange dash-dot-dot). 
			$\Gamma_{sf}/ \Delta_0 = 10^{-3}$ and $0.3$ for left
			and right panels, respectively. We have set
			the incoherent broadening $\delta=10^{-3}\Delta_{0}$. The Seebeck coefficient
			increases with increasing the magnetic field, approaches a
			maximum for a specific field, and eventually drops to zero when
			the exchange field has destroyed the
			superconductivity. The shaded region depicts the exchange fields above the CC limit. Thus, the plotted results correspond to the metastable superconducting state, while the true stable state is normal and corresponds to vanishing thermoelectric response.} 
		\label{fig:Thermopower1}
	\end{centering}
\end{figure}

The Eilenberger equation of motion \cite{Eilenberger1968} for a homogeneous 
spin-split superconductor is given by
\begin{align}
    [iE\hat{\tau}_{3}-\hat{\Delta}-iH_{ex}\sigma\hat{\tau}_{3}-\hat{\Sigma}_{sf},\hat{g}^{R/A}]=0
    \, , \label{eqn_Eilenberger}
\end{align}
where we have set $k_{B}=\hbar=1$. $\hat{g}^{R/A}$ is the retarded
(advanced) Green's function and is obtained by replacing
the quasiparticle energy $E$ by $[E\pm i\delta]$, respectively. The incoherent
broadening $\delta$ or the so-called `Dynes' parameter \cite{Dynes1978}, parametrizes
inelastic scattering and conserves the analytic structure of the 
Green's functions \cite{Belzig1999}. Here, $\hat{\Delta}$ is the superconducting 
pair-potential matrix given as $\hat{\Delta}=\Delta\hat{\tau}_1$, with $\hat{\tau}_i$ 
the Pauli matrices in Nambu space. For a homogeneous superconducting state, 
$\Delta$ may be chosen as real and corresponds to the superconducting gap. 
The gap at zero temperature, exchange field, and spin-flip scattering rate is
denoted by $\Delta_{0}$. The exchange field $H_{ex}$ acts on the
electron spins and splits the DOS for spin-up and spin-down electrons
\cite{Buzdin2005,Golubov2004,Bergeret2005}. $\sigma=\{\pm 1\}$ is a
spin label referring to the up and down orientations of the
spin ($\uparrow/\downarrow$) with respect to the external
spin-splitting field. Since we neglect inhomogeneities, 
gradient-terms do not appear in Eq. \eqref{eqn_Eilenberger}. The self-energy 
corresponding to spin-flip scattering via magnetic impurities~\cite{Kopnin2001,Morten2004,Morten2005,Bergeret2005,Bennemann2008} 
corresponds to 
$\hat{\Sigma}_{sf}=(1/2\tau_{sf})\hat{\tau}_3\hat{g}^{R/A}\hat{\tau}_3$. 
Alternatively, such a self-energy can originate from a magnetic interface 
with strong spin-dependent scattering \cite{Cottet2009,Eschrig2015,Belzig2017} 
thereby influencing the DOS \cite{Strambini2017,Ouassou2017}. 
The spin-flip scattering time $\tau_{sf}$ is the average time between changes 
of the spin state of an electron \cite{Kopnin2001}. 

We employ the following $\vartheta$ parametrization for the quasiclassical 
Green's function matrix
\begin{align}
 \hat{g}^R_\pm = \cos \vartheta_\pm \hat{\tau}_3 + \sin \vartheta_\pm \hat{\tau}_1
 = \begin{pmatrix} \cos \vartheta_\pm & \sin \vartheta_\pm  \\ \sin \vartheta_\pm
 & -\cos \vartheta_\pm  \end{pmatrix}, \label{eqn_parametrization}
\end{align}
which automatically ensures the normalization
$\hat{g}^{R}_\pm\hat{g}^{R}_\pm=\hat{1}$. In these expressions
$\cos\vartheta_\pm$ and $\sin\vartheta_\pm$
are normal and anomalous Green's functions, respectively,  and
$\vartheta_\pm$ is, in general, a complex quantity. Using equations
\eqref{eqn_Eilenberger} and \eqref{eqn_parametrization}, the equation of
motion simplifies to a set of nonlinear equations for up and down-spins as  
\begin{align}
 -2\big((E+i\delta) \pm H_{ex}\big)\sin\vartheta_\pm + 2i\Delta\cos\vartheta_\pm
 -i\Gamma_{sf}\sin2\vartheta_\pm = 0 \, , \label{eqn_setofnonlinear}
\end{align}
with $\Gamma_{sf}=1/\tau_{sf}$ being the spin-flip scattering
rate. The superconducting pair potential is determined self-consistently
from the imaginary component of the anomalous part (pair amplitude)
of the retarded Green's function: 
\begin{align}
    \Delta &= \frac{\lambda}{2}\int_{0}^{\Omega_\textrm{c}}dE\tanh\frac{E}{2k_{B}T}
    \Im[\sin\vartheta_{+}+\sin\vartheta_{-}]  \,, 
\end{align}
where $\lambda$ is the dimensionless Bardeen-Cooper-Schrieffer (BCS)
\cite{BCS1957} interaction constant. $\Omega_\textrm{c}$ is a suitably 
chosen cut-off.
Once we have an expression for the self-consistent
superconducting pair potential, we can calculate the DOS from the real
part of the normal component of the retarded Green's function. Within
this framework, the DOS for spin up ($\uparrow$) and down
($\downarrow$) are 
\begin{align}
    \label{eqn_updownDoS}
    N_{\uparrow}(E)&= N(0)\Re[\cos\vartheta_{+}]
    \,, \notag
    \\
    N_{\downarrow}(E)&= N(0)\Re[\cos\vartheta_{-}]
    \, . 
\end{align}
with $N(0)$ the DOS in the normal state at the Fermi level.
\begin{figure}[t!]
	\begin{centering}
		\includegraphics[width=0.8\textwidth]{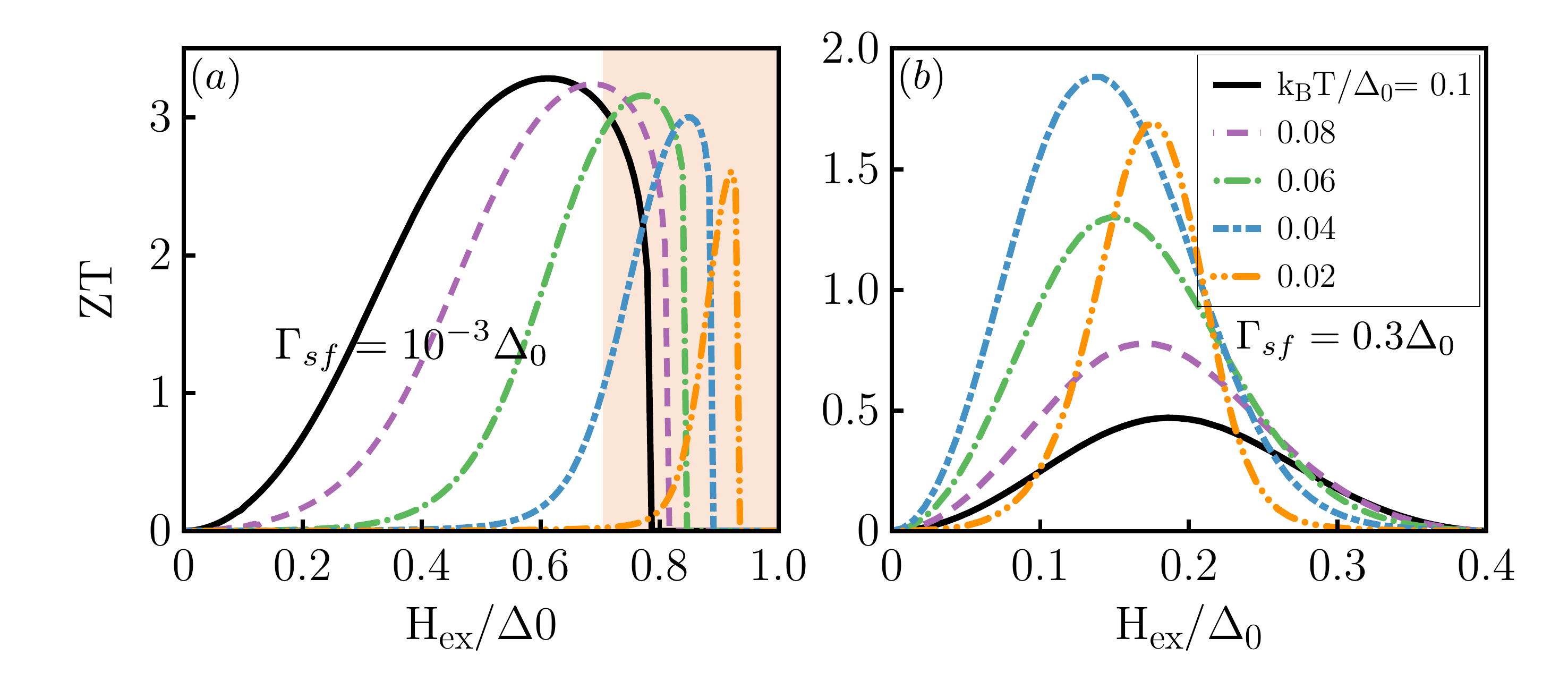}
		\caption{(Color online). Self-consistently determined thermoelectric
			figure of merit $ZT$ as a function of the normalized
			spin-splitting field $H_{ex}/\Delta_{0}$. The parameter 
			values employed are $\Gamma_{sf}/ \Delta_0 = 10^{-3}$ and $0.3$ for left
				and right panels, respectively. We have set polarization of the interface conductance to $P=0.9$
				and the incoherent broadening is $\delta=10^{-3} \Delta_{0}$. The shaded region represents fields larger than the CC limit and the curves displayed pertain to the metastable superconducting state.} 
		\label{fig:ZT}
	\end{centering}
\end{figure}

Our approach outlined above determines the relevant properties of the superconducting state via the self-consistency equation, and then employs the knowledge of these equilibrium properties to evaluate the system response. However, this approach does not guarantee that the equilibrium state thus obtained is lowest in energy, and hence, the `true' ground state. For a superconductor in a magnetic field, the superconducting 
	ground state, disregarding spin-flip scattering, is stable, i.e. lowest in energy, only below the so-called `Chandrasekhar-Clogston' (CC) limit $H_{ex}=\Delta_0/\sqrt{2}$ \cite{CC1962}, which is lower than the 
	Pauli limit $H_{ex}=\Delta_0$ captured by our approach. Hence, our calculations pertain to the superconducting solution of the self-consistency equation, which becomes metastable in certain range of exchange splitting, i.e. its energy is larger than the system being in normal state. The true ground state in this range is the normal state, the thermoelectric response in which is vanishingly small. We determine this range from the literature \cite{Fulde1973,Bergeret2017} and mark it via shaded regions in Figs. \ref{fig:Thermopower1}, \ref{fig:ZT} and \ref{fig:s_max}.

\begin{figure}[th]
	\begin{centering}
		\includegraphics[width=0.82\textwidth]{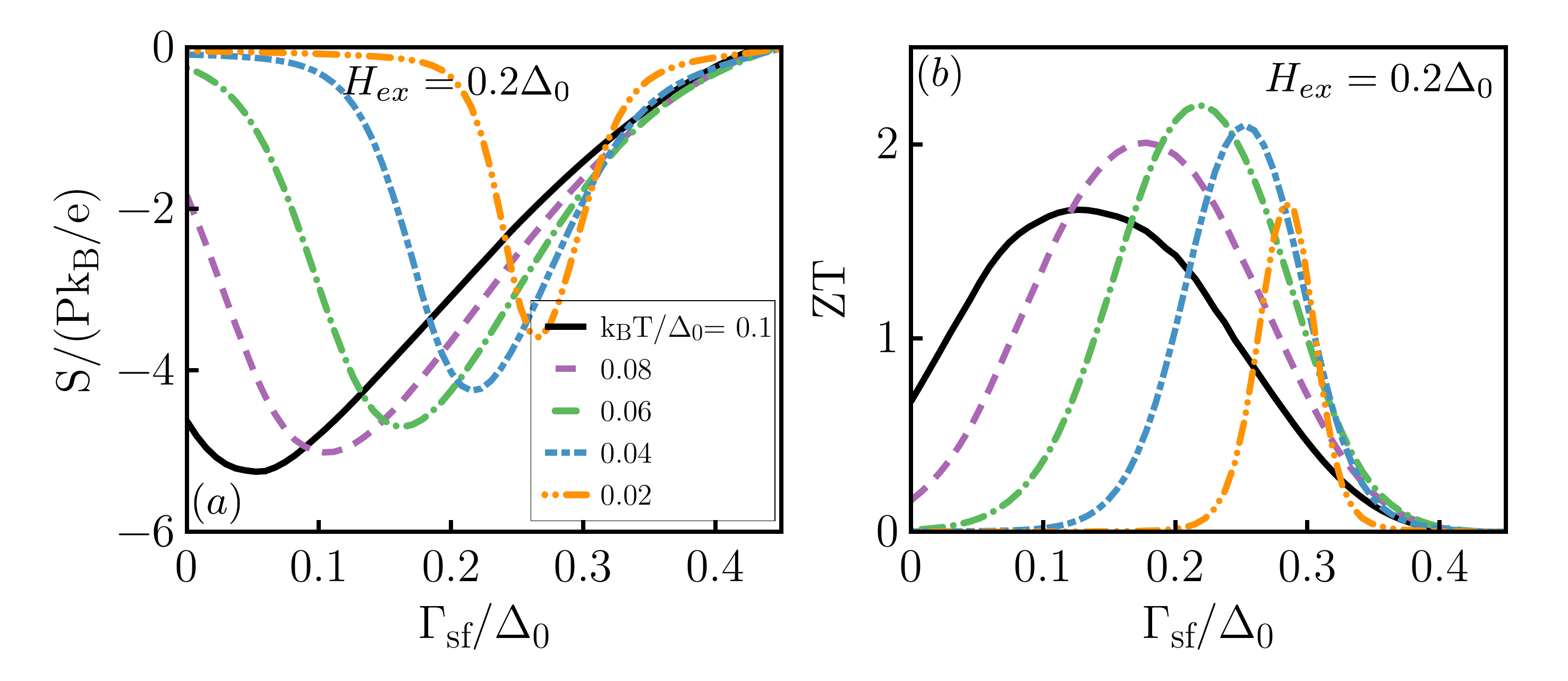}
		\caption{(Color online). Spin-flip scattering rate ($\Gamma_{sf}/\Delta_0$) 
			dependence of the self-consistently determined normalized thermopower 
			$S/(Pk_{B}/e)$ (left panel) and thermoelectric figure of merit 
			$ZT$ (right panel) for $H_{ex}/\Delta_{0}=0.2$ at different 
			temperatures. Polarization of the interface conductance and incoherent broadening are assumed
		    $P=0.9$ and $\delta=10^{-3}\Delta_{0}$.}
		\label{fig:Thermopower&ZT}
	\end{centering}
\end{figure}

\section{Results and discussions} \label{sec:results}

In Fig. \ref{fig:DoS}, we show the total [$N_0(E)$] and 
spin-polarized [$N_z(E)$] DOS obtained within a self-consistent 
evaluation for different exchange fields and a finite spin-flip scattering rate as described above. For the total DOS, the gap slowly closes as the spin-splitting is increased, and via the gapless state, eventually results in 
the destruction of the superconducting state. The spin-polarized DOS vanishes 
for no spin-splitting as well as at-large spin-splittings, when the superconducting 
state is destroyed. Hence, we see that the optimal situation for the TE 
coefficient $\alpha$, corresponding to the largest $N_{z}(E)$ [Eqs. (\ref{eqn_thermocoefficients})], 
is achieved for a certain value of the spin-splitting.

Employing the obtained DOS and Eqs. \eqref{eqn_thermocoefficients}, we 
evaluate the Seebeck coefficient and plot it vs. spin-splitting in Fig. \ref{fig:Thermopower1}. 
As expected from our earlier discussion on spin-polarized DOS, the Seebeck 
coefficient vanishes at zero and large spin-splittings yielding a maximum in 
between. We see that the value of spin-splitting at which the maximal Seebeck coefficient 
is achieved varies strongly with the temperature. Remarkably, in the absence of spin-flip scattering (left panel of Fig.~\ref{fig:Thermopower1}) at the lowest temperatures, the Seebeck coefficient is drastically reduced due to the CC restriction on the field range.

Now comparing the cases of small and large spin-flip rate in Fig. \ref{fig:Thermopower1}, 
we observe that a large spin-flip rate reduces the temperature sensitivity of the 
spin-splitting ($H_{\mathrm{max}}$) corresponding to a maximal Seebeck effect. Overall, putting the CC physics aside for the moment, $H_{\mathrm{max}}$ lowers with an increasing spin-flip rate, 
but this lowering does not have a one-to-one correspondence with the reduction of 
the superconducting gap, as evidenced by the relative insensitivity of $H_{\mathrm{max}}$ 
to the temperature. The TE performance of the device is particularly 
insensitive to the temperature for large spin-flip, as long as 
the system is not in the gapless state. Figure \ref{fig:ZT} presents the analogous 
results for ZT corroborating the similar dependence of TE efficiency on 
the relevant physical variables. We emphasize that ZT values large than 1 can only be 
achieved in the low-temperature and low-exchange field regime by introducing a 
spin-flip rate as shown in the right panel of Fig.~\ref{fig:ZT}. In the absence of spin-flip scattering, ZT is dramatically reduced at very low temperatures $T\lesssim 0.05\Delta_0$ due to the instability of the superconducting state in the CC regime.

We now focus on the dependence of the TE response on the 
spin-flip rate. To this end, we plot the Seebeck coefficient (left panel) and 
ZT (right panel) vs. spin-flip rate at a fixed value of spin-splitting in Fig.~\ref{fig:Thermopower&ZT}. 
 The Seebeck coefficient 
exhibits an initial enhancement with the spin-flip rate followed by an eventual reduction.  This behavior may be understood in terms of spin-flip mediated 
lowering of the superconducting gap. The first order effect of spin-flip in this regime is to 
reduce the superconducting gap thereby shifting the quasiparticle dynamics to lower energies 
without significantly damaging the superconducting state. At larger spin-flip rates, the Cooper 
pair depairing effect dominates the modification of the system TE response. ZT essentially mirrors the dependence of Seebeck coefficient with minor differences.

\begin{figure}[t]
  \begin{centering}
      \includegraphics[width=0.75\textwidth]{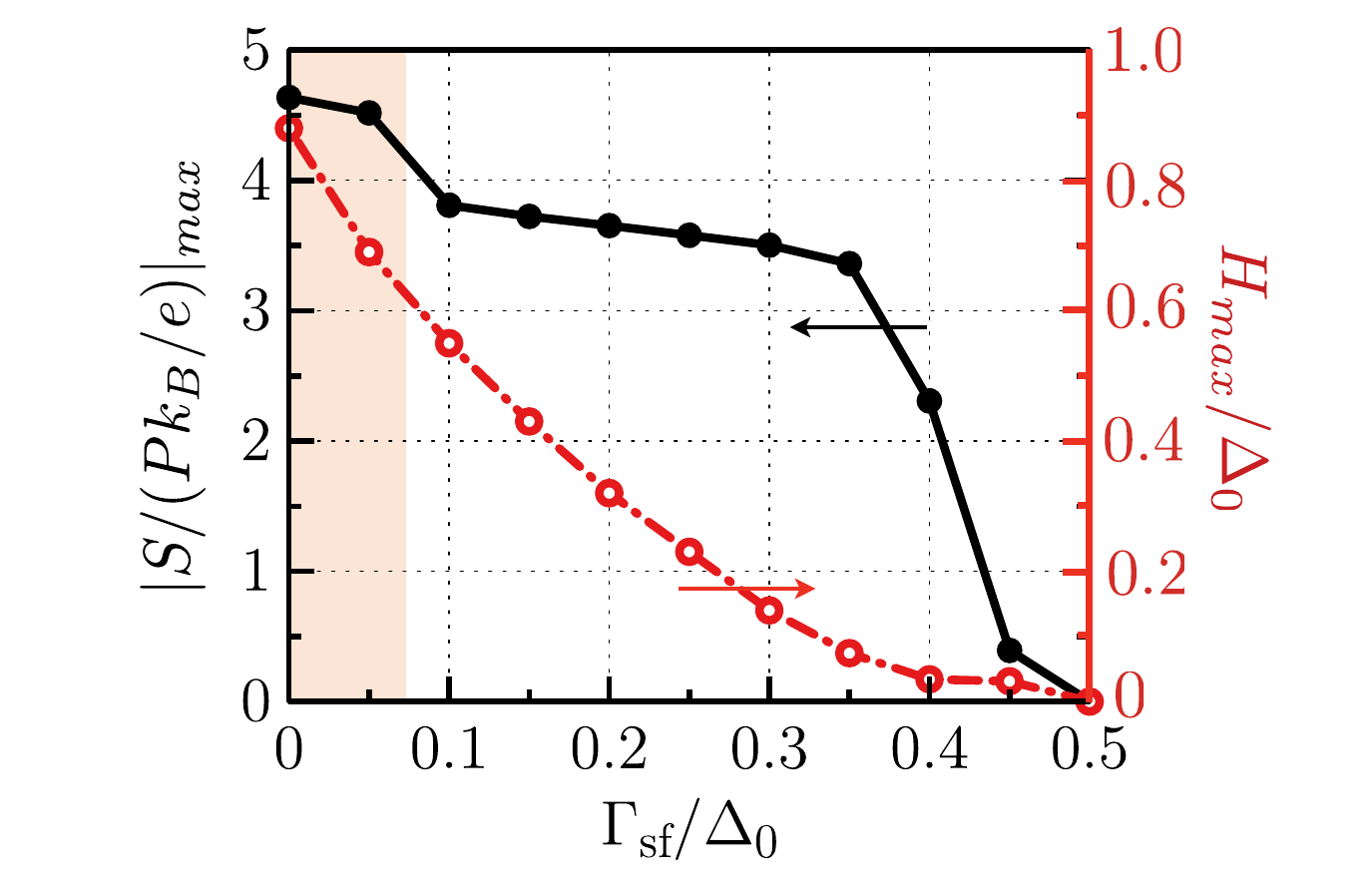}
       \caption{ (Color online). Maximal thermopower $\vert S/(Pk_{B}/e)\vert_{max}$ 
       	vs spin-flip rate $\Gamma_{sf}$ at $k_BT/\Delta_{0}=0.02$ 
       	[the orange curves in Fig. \ref{fig:Thermopower1}]. For a small 
        incoherent broadening $\delta=10^{-3}\Delta_0$, the maximum Seebeck coefficient for medium spin-flip scattering rates is relatively insensitive to the spin-flip rate. The shaded region represents the destruction of the superconducting state due to CC physics. The upper spin-flip rate limit of the shaded region has not been calculated exactly and represents an approximate value. At small spin-flip rates, the true Seebeck-coefficient vanishes due to the required exchange field being above the CC limit. On the other hand, large spin-flip rates destroy superconductivity via the usual depairing channel.} 
      \label{fig:s_max}
    \end{centering}
\end{figure}

We finally study the dependence of the maximum Seebeck coefficient (as a function 
of spin-splitting) on the spin-flip rate in Fig.~\ref{fig:s_max}. The maximum 
Seebeck coefficient is relatively insensitive to the spin-flip. However, the value of 
spin-splitting (not shown explicitly) at which this maximum value is achieved decreases 
monotonically with the spin-flip rate.

Based on the analysis presented above, we are now in a position to summarise 
the effect of spin-flip on the TE response of the system under investigation. 
At small values of the spin-flip scattering rate, its predominant effect is reducing the superconducting gap. 
This provides quasiparticles at lower energies and results in enhancement of Seebeck 
effect at low values of the spin-splitting exchange field. After the superconductor enters the gapless regime, 
there is no such gain in further increasing the spin-flip. On the contrary, the depairing 
effect only tends to make the DOS flatter resulting in a decreasing TE response 
with increasing spin-flip. We note that spin-orbit scattering will mix the spin-split energy bands \cite{Fulde1973} and hence reduce the overall TE performance. Thus, the optimal value of spin-flip is around the point when the 
superconductor is about to enter the gapless regime. 

\section{Conclusions}\label{sec:conclusions}

We have investigated the thermoelectric response of a
tunnel structure formed by a thin insulating layer sandwiched between
a spin-split superconducting film and a ferromagnetic metal. We take
into account the important roles of the spin-flip scattering and Chandrasekhar-Clogston (CC) paramagnetic limit \cite{CC1962,Fulde1973}. The relevant physical quantities have been obtained
self-consistently within the quasiclassical Green's function technique. 
We have shown that increasing the spin-flip scattering rate leads to a 
strong enhancement of the thermoelectric performance in the low-field 
and low-temperature regime. Specifically, the maximum of the thermopower 
exceeds $k_{B}/e$ by a factor of $\approx 5$ down to temperatures of the order of 
$k_{B}T/\Delta_0=0.02$ and has a nonmonotonic behavior with respect 
to the spin-splitting and the spin-flip scattering rate in the superconductor. 
We also predict a sizeable thermoelectric figure of merit of $ZT\gtrsim 1$ in 
the low-temperature regime. Although the spin-flip rate does affect the maximum of the thermopower 
only weakly, it shifts the peak from a higher to a lower field. This appears to be of crucial importance since we find that the high fields required in the absence of spin-flip scattering are larger than the CC limit, and thus do not support superconductivity. Our results constitute a promising prediction for reducing the necessary spin-splitting 
in these structures, which might also be useful to avoid other detrimental effects 
related to external magnetic fields, magnetization control or the CC-limited ultra-low-temperature degradation of the TE response.
\section*{Acknowledgments}\label{sec:acknow}

We gratefully acknowledge discussions with D. Beckmann and thank 
A. J. Pearce for proofreading the manuscript. This work was financially 
supported by the DFG through SPP 1538 Spincaloric Transport and the 
Alexander von Humboldt Foundation.
\section*{References}

\bibliographystyle{phaip}

\end{document}